\documentstyle[12pt,twoside,fleqn,espcrc1,epsfig]{article}

\newcommand{\AmS}{{\protect\the\textfont2
  A\kern-.1667em\lower.5ex\hbox{M}\kern-.125emS}}

\title{Kaon photoproduction on the nucleon: overview of some
       applications}

\author{T. Mart\address{Jurusan Fisika, FMIPA, Universitas Indonesia,
        Depok 16424, Indonesia}\thanks{Supported in part by the 
        University Research for Graduate Education
        (URGE) grant.}, S. Sumowidagdo$^{\rm a*}$, 
        D. Kusno$^{\rm a*}$, 
        C. Bennhold\address{Center for Nuclear Studies, Department 
        of Physics, The George Washington University, \\ 
        ~\,Washington, D.C. 20052, USA}\thanks{Supported in part by
        US DOE with grant no. DE-FG02-95ER-40907},
        and H. Haberzettl$^{\rm b\dagger}$}

\begin{document}
% typeset front matter
\maketitle

\begin{abstract}
  Some applications of the elementary kaon photoproduction process
  are discussed: the investigation of missing resonances 
  in the $p(\gamma,K^+)\Lambda$ channel, the 
  role of the $P_{13}(1720)$ state in the $p(\gamma,K^0)\Sigma^+$ channel, 
  and the calculation of the Gerasimov-Drell-Hearn sum rule. For the latter,
  we present an extension of our previous study to higher energies.
\end{abstract}

\section{INTRODUCTION}
The electromagnetic production of kaons provides an important source of 
information on hadronic physics with strangeness. Due to the rather weak
coupling at the electromagnetic vertex, the reaction operator 
is quite simple and,
therefore, the use a of single channel analysis may be quite sufficient to 
describe the process.  This process can be used to
simultaneously study kaon-hyperon-nucleon coupling constants, 
isospin symmetry, hadronic form factors, baryon and meson resonances, 
and contributions of 
kaon-hyperon final states to the magnetic moment of the nucleon. In this
report we will focus only on the two last topics.

\section{THE ELEMENTARY PROCESS WITH SOME APPLICATIONS}
For a detailed discussion of the elementary operator we refer to 
Ref.\,\cite{fxlee99}. 
The background part of the operator consists of the standard Born terms 
along with the $K^*$(892) and $K_1$(1270) vector meson poles in 
the $t$-channel. The resonance part of the $K\Lambda$ 
operator includes 
the $S_{11}$(1650), $P_{11}$(1710), and $P_{13}(1720)$ resonances. 
We also include hadronic form factors in hadronic vertices by employing 
the gauge method of Haberzettl \cite{hbmf98} which leads to
an excellent agreement between experimental data and model calculations.

\subsection{Investigation of Nucleon Resonances}
A brief inspection of the Particle Data Table reveals that less than 40\% 
of the nucleon resonances predicted by constituent quark models are observed in 
$\pi N\to \pi N$ scattering experiments. Quark model studies have 
suggested that those "missing" resonances may couple to 
other channels, such as the $K \Lambda$ and $K \Sigma$ channels. 
Stimulated by the new $p(\gamma, K^+)\Lambda$ {\footnotesize SAPHIR} 
total cross section data \cite{saphir98}, which suggest a structure 
around $W = 1900$ MeV, we start the investigation by using our isobar
model.
\begin{figure}[!t]
\begin{minipage}[htb]{77mm}
\epsfig{file=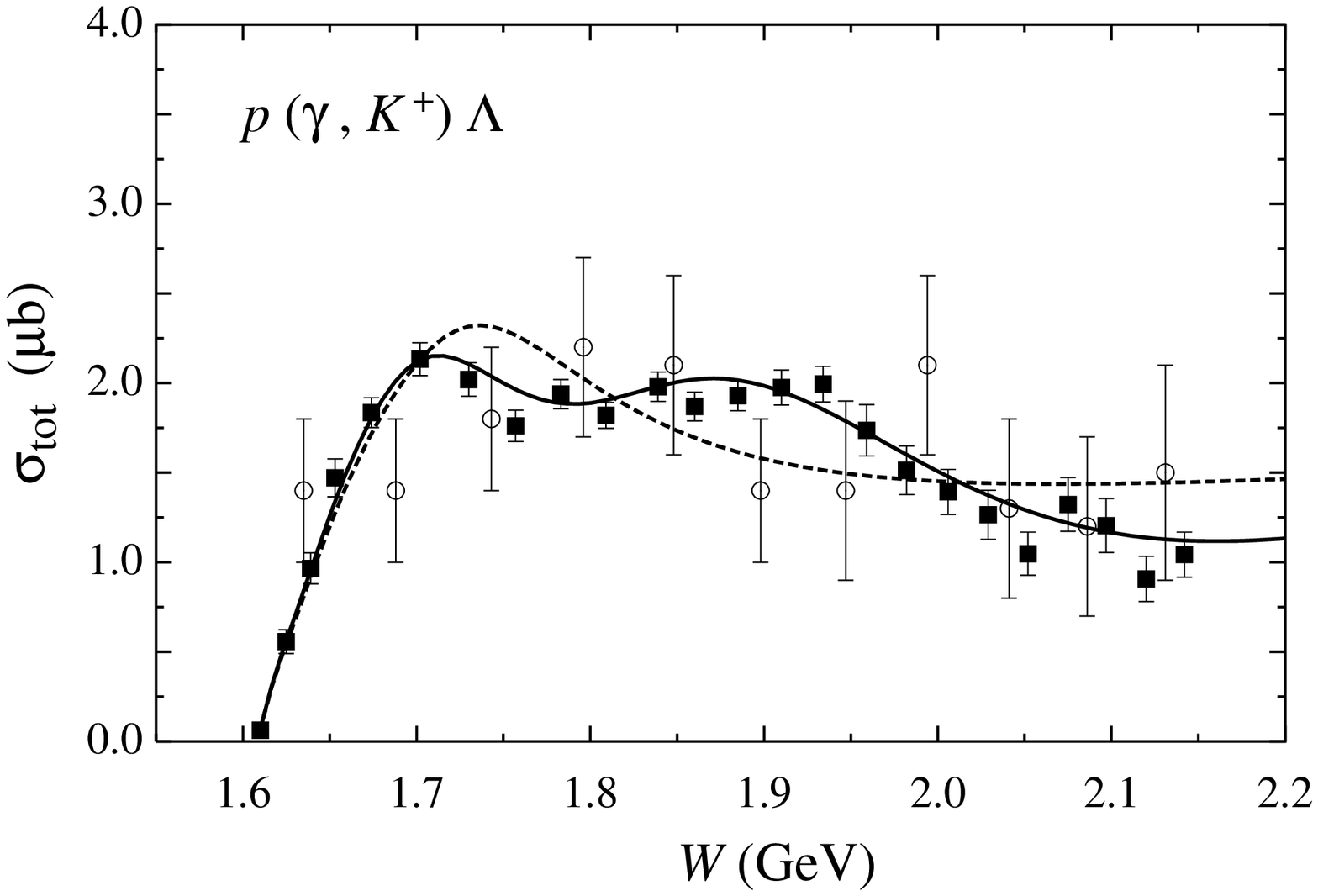, width=77mm}
\vspace{-15mm}
\caption{Total cross section for $p(\gamma,K^+)\Lambda$. 
        The dashed line shows the model without 
        $D_{13}(1895)$ resonance, while the solid line is obtained by including 
        $D_{13}(1895)$ resonance. Experimental data 
        from {\footnotesize SAPHIR} are shown by the 
        solid squares \protect\cite{saphir98}. 
        Old data are displayed by the open
        circles \protect\cite{old_data}.}
\label{fig:missing}
\end{minipage}
\hspace{\fill}
\begin{minipage}[htb]{77mm}
\epsfig{file=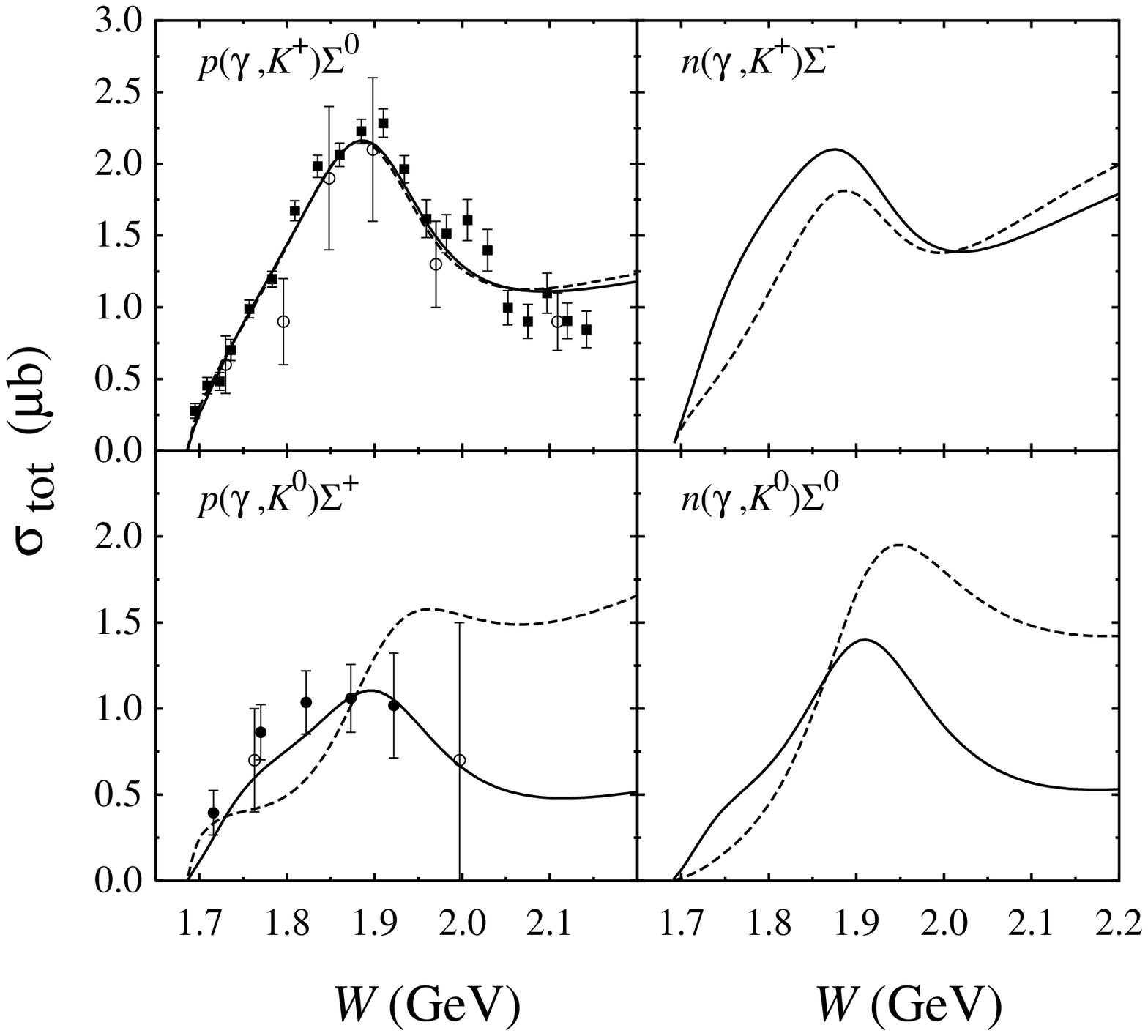, width=80mm}
\vspace{-15mm}
\caption{Total cross sections for $K\Sigma$ photoproduction.
        The dashed (solid) lines show the model  
         without (with) the $P_{13}(1720)$ 
         resonance \protect\cite{mart_p13}. The new 
         {\footnotesize SAPHIR} data are shown by the solid circles
         \protect\cite{saphir99}.}
\label{fig:p13role}
\end{minipage}
\vspace{-0.6cm}
\end{figure}
As shown in Fig.\,\ref{fig:missing}, our previous model 
cannot reproduce the total cross section. 
The constituent quark model of Capstick and Roberts \cite{capstick98} 
predicts many new states around 1900 MeV. However, only a few of them have 
been calculated to have a significant $K \Lambda$ decay width 
\cite{capstick98}. We have performed fits for each 
of the possible states, allowing the fit to determine the mass, width 
and coupling constants  of the resonance. We found that only 
in the case of the $D_{13}$(1895) state couplings are obtained that
are in remarkable agreement \cite{terry2000} with recent quark model predictions.
The result is shown in Fig.\,\ref{fig:missing}, where 
without this resonance the model shows only one peak
near threshold, while inclusion of the new resonance leads
to a second peak at $W$ slightly below 1900 MeV,
in accordance with the new {\small SAPHIR} data.

In the case of $K^0\Sigma^+$ photoproduction we find 
that a better agreement with experimental data can be
achieved by including the $P_{13}(1720)$ resonance. 
As shown in Fig.\,\ref{fig:p13role}
the inclusion of this state does not influence the $K^+\Sigma^0$ 
channel appreciably, in contrast to the other three $K\Sigma$
channels, where the prominent effect is found in the $K^0\Sigma^+$
channel. This is due both to the isospin coefficient which enhances
the $P_{13}(1720)\to K^0\Sigma^+$ decay over the
$P_{13}(1720)\to K^+\Sigma^0$ coupling and the very different structure of the Born terms
in $K^0$ photoproduction.
%Our result 
%implies that the $P_{13}(1720)\to K^0\Sigma^+$ decay mode 
%is more likely to occur rather than the  $P_{13}(1720)\to K^+\Sigma^0$ mode.
The extracted fractional decay width is found to be 
consistent with the prediction of the quark model, whose magnitude is 
almost one order smaller than the value given by Particle Data 
Group \cite{mart_p13}.

\subsection{Gerasimov-Drell-Hearn Sum Rule}

The Gerasimov-Drell-Hearn (GDH) sum rule \cite{gdh} relates the anomalous
magnetic moment of the nucleon $\kappa_N$ to its excitation spectrum in the
resonance  region by an integral.  Previous analyses on pion
photoproduction multipole amplitudes show a discrepancy between theoretical
prediction and indirect experimental investigations of the GDH sum 
rule \cite{karliner}.  Here we calculate the contribution of kaon 
photoproduction processes to the GDH sum rule by using our isobar model.

In our previous calculation \cite{suharyo99}, we have fixed the upper 
limit of integration at 2.2 GeV. Exact formulation was obtained by 
integrating the structure function $\sigma_{\rm TT'}$, and an approximation 
of the upper bound of integral is achieved by calculating total cross 
section $\sigma_{\rm T}$.  
We obtained a value of $\kappa^2_{p} (K) = -0.063$ and
$\kappa^2_{n} (K) = 0.031$.  Note that although this is  
relatively small, the result is consistent with the Karliner's 
work \cite{karliner}.

\begin{figure}[!t]
\begin{minipage}[htb]{100mm}
    \epsfig{file=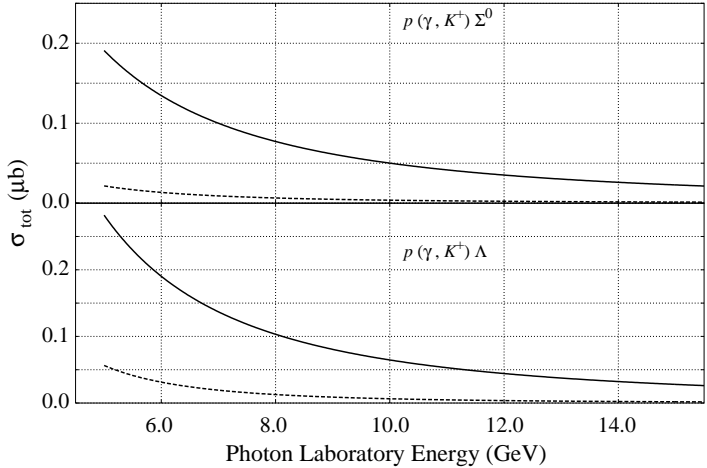,height=95mm,angle=-90}
\end{minipage}
\hspace{\fill}
\begin{minipage}[htb]{50mm}
%\vspace{-1cm}
    \caption{\label{fig:glv}Total cross sections $\sigma_{\rm T}$ (solid lines)
      and $-\sigma_{\rm TT'}$ (dashed lines) for kaon photoproduction on
      the proton at high photon energy.  The elementary model is due to 
      Ref. \cite{glv}.}
\label{fig:gdh}
\end{minipage}
\vspace{-5mm}
\end{figure}

In order to improve our GDH calculation we increase the upper limit of
integration to higher energies.  For the present, we use the Regge model 
given in Ref.\,\cite{glv} to calculate the cross sections in
the energy region between 5 and 15 GeV.  The result is shown in 
Fig.\,\ref{fig:glv}, where it is clear that contributions from higher
energies are very small. Nevertheless, the important message here is the
convergence of $\sigma_{\rm TT'}$ which indicates the convergence of the
integral in the case of kaon.
In the future, we will improve our model by reggeizing the $t-$channel 
intermediate state.  Some preliminary
results have been reported in Ref. \cite{kek}.

%\section{CONCLUSION}
%In conclusion


\begin{thebibliography}{9}
\bibitem{fxlee99} F. X. Lee, T. Mart, C. Bennhold, and L. E. Wright, 
                  nucl-th/9907119 (1999);\\
                  T. Mart, C. Bennhold, and C. E. Hyde-Wright, Phys. Rev. C
                  51 (1995) R1074.
\bibitem{hbmf98} H. Haberzettl, C. Bennhold, T. Mart, and T. Feuster, Phys.
                 Rev. C 58 (1998) R40. 
\bibitem{capstick98} S. Capstick and W. Roberts, Phys. Rev. D 58
                     (1998) 074011.
\bibitem{terry2000} T. Mart and C. Bennhold, Phys. Rev. C 61 (2000) 012201(R).
\bibitem{saphir98} M. Q. Tran {\it et al}., Phys. Lett. B 445 (1998) 20.
\bibitem{saphir99} S. Goers {\it et al}., Phys. Lett. B 464 (1999) 331.
\bibitem{old_data} ABBHHM Collaboration, Phys. Rev. 188 (1969) 2060. 
\bibitem{mart_p13} T. Mart, Phys. Rev. C 62 (2000) 038201.
\bibitem{gdh} S.B. Gerasimov, Yad. Fiz. 2 (1965) 598;
              Sov. J. Nucl. Phys. 2 (1966) 430; \\ S.D. Drell and
              A.C. Hearn, Phys. Rev. Lett. 16 (1966) 908.
\bibitem{karliner} I. Karliner, Phys. Rev. D 7 (1973) 2717.
%                 D. Drechsel and G. Krein, Phys. Rev. D 58
%                 (1998) 116009.
\bibitem{suharyo99} S. Sumowidagdo and T. Mart, Phys. Rev. C 
                    60 (1999) 028201.
\bibitem{glv} M.~Guidal, J.-M. Laget, and M. Vanderhaeghen, 
                 Nucl. Phys. A627 (1997) 645.
\bibitem{kek} T. Mart, S. Sumowidagdo, C. Bennhold, and H. Haberzettl,
  nucl-th/0002036 (2000).% (to be published by Elsevier Science).
\end{thebibliography}
\end{document}